\documentclass[aps,prb,a4paper,preprint,showpacs]{revtex4}
\usepackage{graphicx}
\usepackage{amsmath}
\usepackage{amsfonts}
\usepackage{amssymb}
\usepackage{color}
\usepackage{amsbsy}
\usepackage{bm}
\usepackage{epstopdf}
\usepackage{epsfig}

\begin{document}
\title{Unconventional topological Hall effect in high-topological-number skyrmion crystals}
\author{Rui Zhu\renewcommand{\thefootnote}{*}\footnote{Corresponding author.
Electronic address:
rzhu@scut.edu.cn} and Weinan Liu  }
\address{Department of Physics, South China University of Technology,
Guangzhou 510641, People's Republic of China   }

\begin{abstract}

Skyrmions with the topological number $Q$ equal an integer larger than 1 are called high-topological-number skyrmions or high-$Q$ skyrmions. In this work, we theoretically study the topological Hall effect in square-lattice high-$Q$ skyrmion crystals (SkX) with $Q=2$ and $Q=3$. As a result of the emergent magnetic field, Landau-level-like electronic band structure gives rise to quantized Hall conductivity when the Fermi energy is within the gaps between adjacent single band or multiple bands intertwined. We found that different from conventional ($Q=1$) SkX the Hall quantization number increases by $1/Q$ in average when the elevating Fermi energy crosses each band. We attribute the result to the fact that the Berry phase ${\cal{C}}$ is measured in the momentum space and the topological number of a single skyrmion $Q$ is measured in the real space. The reciprocality does not affect the conventional SkX because $Q=1=1/Q$.

\end{abstract}

\maketitle
\section{Introduction}

Magnetic skyrmion\cite{SkyrmePRSLSA1961, NagaosaNatNano2013, MuhlbauerScience2009, SekiScience2012, MunzerPRB2010, HamamotoPRB2015, GobelPRB2017} is the spin vortex structure in ferromagnets with a nontrivial topological number $Q$ of the two-dimensional classical spin field ${\bf{n}}\left( {\bf{r}} \right)$, with $Q$ equal to the surface integral of the solid angle of ${\bf{n}}\left( {\bf{r}} \right)$, i.e. $Q = \int {q{d^2}{\bf{r}}} $ with $q = \frac{1}{{4\pi }}{\bf{n}} \cdot \left( {\frac{{\partial {\bf{n}}}}{{\partial x}} \times \frac{{\partial {\bf{n}}}}{{\partial y}}} \right)$. Concerning the symmetry of the skyrmion, one can write the spin field of the skyrmion in a general form of ${\bf{n}}\left( {\bf{r}} \right) = \left[ {\sin \Theta \left( r \right)\cos \Phi \left( \varphi  \right),\sin \Theta \left( r \right)\sin \Phi \left( \varphi  \right),\cos \Theta \left( r \right)} \right]$, with $r$ and $\varphi $ the polar coordinate in the real space, $\Theta $ and $\Phi $ polar and azimuthal angles of the local spin. The topological number of a single skyrmion can be obtained as
\begin{equation}
Q = \frac{1}{{4\pi }}\left[ {\cos \Theta \left( r \right)} \right]_{r =  \infty  }^{r = 0}\left[ {\Phi \left( \varphi  \right)} \right]_{\varphi  = 0}^{\varphi  = 2\pi }.
\end{equation}
When the center spin points upward and the edge spin points downward, or vice versa, we can have the polarity (vorticity) of the skyrmion $ m= \left[ {\cos \Theta \left( r \right)} \right]_{r = \infty }^{r = 0} =  \pm 1$. When the skyrmion whirls in the pattern of $\Phi \left( \varphi  \right) = \xi \varphi  + \gamma $, the topological number is explicitly expressed by $Q = m\xi $ and $\gamma $ determines the helicity of the skyrmion. Conventional magnetic skyrmions have $Q=\pm 1$, with the sign distinguishing skyrmions and antiskyrmions. By varying the whirling period $\xi$ of the azimuthal angle of the local spin $\Phi$, high-topological-number skyrmions with $Q>1$ are theoretically predicted\cite{XichaoZhangPRB2016, HayamiPRB2019, OzawaPRL2017, GalkinaPRB2009}. When identical skymions form a spatially periodic array, we obtain a skyrmion crystal (SkX). Top view of the square-lattice high-topological-umber SkX with $Q=2$ and that of the conventional SkX with $Q=1$ are shown in Figs. 1 and 2, respectively.

The conventional SkX with $Q=1$ spin vortices forming two-dimensional hexagonal, triangular, or square crystal structures were recently discovered in magnetic metal alloys, insulating multiferroic oxides, and the doped semiconductors\cite{NagaosaNatNano2013, MuhlbauerScience2009, SekiScience2012, MunzerPRB2010, HamamotoPRB2015, XZhangSR2015, XXingPRB2016, YZhouNatCommun2014, SampaioNatNano2013, RommingScience2013, TchoePRB2012, IwasakiNatNano2013, SZLinPRL2014, HalsPRB2014, JLiNatCommun2014, EverschorPRB2012, HeinzeNatPhys2011}. Their spin structures can be detected by neutron scattering\cite{MuhlbauerScience2009} and Lorentz transmission electron microscopy\cite{SekiScience2012}. The Hall effect measurements in the SkX metals establish the physics of emergent electrodynamics\cite{NeubauerPRL2009, SchulzNatPhys2012, GobelPRB2019}. Since its discovery, the SkX has attract intensive interest due to its fundamental meaning and potential application in topological computers\cite{NagaosaNatNano2013, SchulzNatPhys2012}. The generation, deleting, and dynamics of isolated skyrmion and SkX have been investigated by theory, numerical simulation, and experiments\cite{IwasakiNatNano2013, YZhouNatCommun2014, SampaioNatNano2013, RommingScience2013, JLiNatCommun2014}. Among these, the topological Hall effect of the SkX resulting from the emergent magnetic field of the nontrivial spin field has attract a lot of attention\cite{HamamotoPRB2015, SchulzNatPhys2012, GobelPRB2017, GobelPRB2019}.

Recently, various works have considered the creation and manipulation of high-topological-number skyrmions or SkX with $Q=2$. Some of the works we happen to come across are: Zhang and \emph{et al}. found that such magnetic skyrmions can be created and stabilized in the chiral magnet with Dzyaloshinskii-Moriya interaction by applying vertical spin-polarized current nonequilibriumly subsisting on a balance between the energy injection from the current and the energy dissipation by the Gilbert damping\cite{XichaoZhangPRB2016}; Ozawa and \emph{et al}. found that the SkX with $Q=2$ can be stabilized in itinerant magnets at a zero magnetic field by investigation of the Kondo lattice model on a triangular lattice\cite{OzawaPRL2017}; Hayami and Motome demonstrated the robustness of the $Q=2$ SkX against single-ion anisotropy on a triangular lattice by numerical simulation of the Kondo lattice model\cite{HayamiPRB2019}; Even earlier Galkina and \emph{et al}. found that skyrmions with $Q=2$ can exist in a classical two-dimensional Heisenberg model of a ferromagnet with uniaxial anisotropy by a variational approach\cite{GalkinaPRB2009}.

This work was inspired by the seminal work of Hamamoto and Nagaosa in 2015 focusing on the topological Hall effect of conventional square-lattice SkX in the strong-Hund's-coupling limit. They found quantized Hall conductivity with the Berry phase of each band contribute a unity $e^2/h$ to the conductivity. We extended the model to high-topological-number SkX with the topological number of a single skyrmion $Q=2$ and $3$, respectively. The technique roots in the exact diagonalization of the tight-binding model on a lattice with a giant unit cell and sublattice-dependent hopping energy. Before the result was obtained, we suppose that the Hall conductivity should quantize at $Q$ multiples of $ e^2/h$ with each band contributing a topological number of $Q$ following that of a single skyrmion. However, we obtain the surprising result that the Berry phase of a single band is as well quantized with the quantization value varying between $0$ to $\pm 3$ and every sequential $Q$ bands form a group, which bears a total Berry phase of $1$. In this way, the Berry phase of each band averages to be $1/Q$. After the work is finished, we noticed the work by G\"{o}bel, \emph{et al}. in 2017 focusing on the topological Hall effect of a triangular-lattice conventional SkX. They found that the Hall conductivity quantized at even-integer times of $e^2/h$ and attributed the result to the topology of the crystal. This is because the topological Hall conductivity is a combined result of the topology of a single skyrmion and that of the crystal. After this work is finished, we also noticed that recently the topological Hall effect beyond the strong coupling regime has been considered and found that with weaker Hund's coupling the Hall conductivity becomes unquantized and varies with the strength of the Hund's exchange\cite{DenisovSR2017, DenisovPRB2018, NakazawaJPSJ2018, NakazawaPRB2019}. We found these findings highly instructive and the Hall effect in high-topological-number SkX traversing the strong and weak coupling regimes would be our future considerations.

  The other parts of the work is organized as follows. Sec. II is the theory and technique. Sec. III is numerical results and discussions. A conclusion is given in Sec. IV.

\section{Model and formalism}

We consider the free-electron system coupled with the
background spin texture ${{\bf{n}}_i}$ by Hund's coupling. ${{\bf{n}}_i}$ is the atomic-lattice-discretized version of the magnetic spin field ${\bf{n}}\left( {\bf{r}} \right)$ introduced in the previous section. Hamiltonian of the electron is given by the double-exchange model\cite{HamamotoPRB2015},
\begin{equation}
H = t\sum\limits_{\left\langle {i,j} \right\rangle } {c_i^\dag {c_j}}  - J\sum\limits_i {{{\bf{n}}_i}c_i^\dag {\bf{\sigma }}{c_i}}  ,
\end{equation}
where ${c_i} = {\left( {{c_{i \uparrow }},{c_{i \downarrow }}} \right)^{\rm{T}}}$ is the two-component annihilation operator at the $i$ site, $c_i^\dag $ is its creation counterpart. $t$ is the hopping integral between nearest-neighbor sites. We assume it the same at all lattice sites. $J$ is the strength of the Hund's coupling between the electron spin and background
spin texture. ${\bf{\sigma }}$ denotes the Pauli matrix.

In the limit that $J \gg t$, the spin of the hopping electron is forced to align parallel to the spin texture. Because there is no other spin-flipping mechanism, hopping can only occur between electrons with parallel spins. We can arrive at a ``tight-binding" model with the effective transfer energy site dependent and equal to $t$ multiplied by the magnitude of the spin overlap\cite{HamamotoPRB2015}. Strength of the spin overlap between sites $i$ and $j$ can be obtained by $\left\langle {{\chi _i}} \right|\left. {{\chi _j}} \right\rangle $ with $\left| {{\chi _i}} \right\rangle $ the wave function of the conduction electron at site $i$ corresponding to the localized spin ${{\bf{n}}_i}$. Using spherical coordinates in the spin space of the electron ${{\bf{n}}_i} = \left( {\sin {\Theta _i}\cos {\Phi _i},\sin {\Theta _i}\sin {\Phi _i},\cos {\Theta _i}} \right)$, we can obtain
\begin{equation}
\left| {{\chi _i}} \right\rangle  = {\left( {\cos \frac{{{\Theta _i}}}{2},{e^{i{\Phi _i}}}\sin \frac{{{\Theta _i}}}{2}} \right)^{\rm{T}}}.
\end{equation}
The effective Hamiltonian can be expressed as\cite{HamamotoPRB2015}
\begin{equation}
H = \sum\limits_{\left\langle {i,j} \right\rangle } {t_{{\rm{eff}}}^{ij}d_i^\dag {d_j}} ,
\label{EffectiveHamiltonian}
\end{equation}
with
\begin{equation}
t_{{\rm{eff}}}^{ij} = t\left\langle {{\chi _i}} \right|\left. {{\chi _j}} \right\rangle  = t\cos \frac{{{\Theta _i}}}{2}\cos \frac{{{\Theta _j}}}{2} + t\sin \frac{{{\Theta _i}}}{2}\sin \frac{{{\Theta _j}}}{2}{e^{ - i\left( {{\Phi _i} - {\Phi _j}} \right)}}.
\label{tEffective}
\end{equation}
Here ${d_i^\dag }$ (${{d_i}}$) is the spinless creation (annihilation) operator at
the $i$ site.

Considering the periodic structure of SkX, Eq. (\ref{EffectiveHamiltonian}) can be taken as a ``tight-binding" model of electrons on a lattice of giant unit cells. Each unit cell corresponds to a single skyrmion. We can rewrite the effective model as
\begin{equation}
H = \sum\limits_{i,\bf{\delta} } {{t^{s,s'}}d_{i,s}^{\dagger} {d_{i + \bf{\delta} ,s'}}} ,
\label{GiantUnitCellHamiltonian}
\end{equation}
where the summation of $i$ goes through the complete atomic lattice, the summation of $\delta$ goes through the four nearest-neighbor sites of the $i$-\emph{th} site, site $i$ locates on the $s$-type sublattice and site $i + \delta$ locates on the $s'$-type sublattice, and ${t^{s,s'}} = t_{{\rm{eff}}}^{ij}$ with $j=i+ \delta$. Schematics of the model is shown in Fig. 3. Four examples of $t^{s,s'}$, i.e., $t^{\rm{BA}}$, $t^{\rm{UA}}$, $t^{\rm{EA}}$, and $t^{\rm{FA}}$ are shown in the figure with the corresponding $\bf{\delta}$ are $-a{{\bf{\hat e}}_x}$, $-a{{\bf{\hat e}}_y}$, $a{{\bf{\hat e}}_x}$, $a{{\bf{\hat e}}_y}$, respectively. There are two lattice constants: one is $a$ measuring the distance between adjacent atoms; the other is $2 \lambda$ measuring the size and interval of the skyrmions. Without loss of generality, radius of the skyrmion $\lambda$ is set to be $2.5a$ and a single skyrmion consists of $5 \times 5 = 25$ atoms. By Fourier transformation of Eq. (\ref{GiantUnitCellHamiltonian}), we can obtain
\begin{equation}
H = \frac{1}{N}\sum\limits_{i,{\bf{\delta }},{{\bf{k}}_1},{{\bf{k}}_2}} {{t^{s,s'}}d_{{{\bf{k}}_2},s}^\dag {d_{{{\bf{k}}_1},s'}}{e^{i{{\bf{k}}_1} \cdot \left( {{{\bf{r}}_i} + {\bf{\delta }}} \right) - i{{\bf{k}}_2} \cdot {{\bf{r}}_i}}}}  = \sum\limits_{{\bf{\delta }},{\bf{k}}} {{t^{s,s'}}d_{{\bf{k}},s}^\dag {d_{{\bf{k}},s'}}{e^{i{\bf{k}} \cdot {\bf{\delta }}}}} ,
\end{equation}
which is diagonal in the $\bf{k}$-space and a $25 \times 25$ matrix in the sublattice space. We consider a background spin texture ${\bf{n}}\left( {\bf{r}} \right)$ made of a high-topological-number square-lattice SkX. Each skyrmion has a nontrivial topological number $Q=2$. The skyrmion profile is well assumed as $\Theta \left( r \right) = \pi \left( {1 - {r \mathord{\left/
 {\vphantom {r \lambda }} \right.
 \kern-\nulldelimiterspace} \lambda }} \right)$ for $r< \lambda$, $\Theta \left( r \right) = 0$ for $r> \lambda$, and $\Phi \left( \varphi  \right) = 2\varphi  + \gamma $. It is obvious from Eqs. (\ref{EffectiveHamiltonian}) and (\ref{tEffective}) that different $\gamma$ makes no difference to the effective Hamiltonian of the conducting electrons. The emergent magnetic field is produced by the spin texture with the total magnetic flux ${{Qh} \mathord{\left/
 {\vphantom {{Qh} e}} \right.
 \kern-\nulldelimiterspace} e}$, which is independent of the skyrmion radius $\lambda$. By exact diagonalization of the $25 \times 25$ matrix at each point in the $\bf{k}$-space, we can obtain the band structure $E_{n\bf{k}}$ (eigenvalues of the matrix) and the electronic states $\left| {n{\bf{k}}} \right\rangle $.

The Chern number of each band is the integral of the Berry curvature over the first Brillouin zone
\begin{equation}
{\cal{C}} = \int {{d^2}k{b_z}\left( {\bf{k}} \right)}
\label{BerryPhase}
\end{equation}
with
\begin{equation}
{b_z}\left( {\bf{k}} \right) = \frac{{ - i}}{{2\pi }}\left[ {{\partial _{{k_x}}}\left\langle {n{\bf{k}}} \right|{\partial _{{k_y}}}\left| {n{\bf{k}}} \right\rangle  - {\partial _{{k_y}}}\left\langle {n{\bf{k}}} \right|{\partial _{{k_x}}}\left| {n{\bf{k}}} \right\rangle } \right].
\label{BerryCurvature}
\end{equation}
The topological Hall conductivity at zero-temperature calculated from the Kubo formula is\cite{HamamotoPRB2015, ThoulessPRL1982}
\begin{equation}
{\sigma _{xy}} = \frac{{{e^2}}}{h}\int_\Omega  {\frac{{ - i}}{{2\pi }}\sum\limits_{{E_n} < {E_F},m \ne n} {\frac{{\left\langle {n{\bf{k}}} \right|\frac{{\partial H}}{{\partial {k_x}}}\left| {m{\bf{k}}} \right\rangle \left\langle {m{\bf{k}}} \right|\frac{{\partial H}}{{\partial {k_y}}}\left| {n{\bf{k}}} \right\rangle  - \left( {n \leftrightarrow m} \right)}}{{{{\left( {{E_{n{\bf{k}}}} - {E_{m{\bf{k}}}}} \right)}^2}}}} d{k_x}d{k_y}} =\frac{e^2}{h}\sum\limits_{{E_n} < {E_F}} {\cal C},
\label{sigma}
\end{equation}
in which $\Omega$ is the first Brillouin zone. The Hall conductivity is equal to the total Berry phase below the Fermi energy in the unit of $e^2/h$. In direct diagonalization of the Hamiltonian, the phase factor of the eigenspinor is not definite, which is also called gauge of the state. Different gauge induces a sign difference in the Berry phase of a particular band. Usually, the gauge sets the $n$-\emph{th} row of the eigenspinor to be unity if one directly calculate the Berry phase from geometry of the eigenstate. However, this procedure is not necessary because the Hall conductivity formula always has the bra and ket of the eigenspinor in pair. One should only be careful to use the same gauge throughout one work.

\section{Results and discussions}

Numerical results of the high-topological-number SkX with $Q=2$ are given in Fig. 1. For comparison, those of the conventional SkX with $Q=1$ are given in Fig. 2. We can see that $n_x$ and $n_y$ of the $Q=2$ SkX have a four-leaf structure and $n_y$ is $n_x$ clockwisely rotated by ${\pi}/4$; $n_x$ and $n_y$ of the $Q=1$ SkX have a double-leaf structure and $n_y$ is $n_x$ clockwisely rotated by ${\pi}/2$. Profile of the spin fields shows the whirling pattern of ${\xi}=2$ and $1$, respectively. Because the polar angle of the local spin $\Theta$ only depends on the polar radius $r$ in the real space with respect to the center of each skyrmion, $n_z$ of $Q=2$ and $Q=1$ skyrmions have identical patterns. The top view of the topological charge density distribution $q$ in the real space of the two types of the SkX is shown in panel (d) of the two figures. It can be seen that different from conventional SkX $q$ of the $Q=2$ SkX has a fine structure in the region near the center of the each skyrmion. This is a demonstration of the topological difference between the high-topological-number SkX and the conventional SkX, which induces the unconventional topological Hall conductivity as discussed below.

Zero-temperature topological Hall conductivity calculated from Eq. (\ref{sigma}) is given in panel (g) of the two figures. Quantized ${\sigma}_{xy}$ in the gap between adjacent bands is a direct result of the nontrivial topology of the band structure of the SkX and the quantization number is equal to the total Berry phase of the bands below the Fermi energy of the conducting electrons. Panels (e) and (f) of the two figures are numerical results of the band structure of the lowest ten bands in the vicinity of the $M$ $({\pi}/{\lambda},{\pi}/{\lambda})$ point in the first Brillouin zone. The Berry phase ${\cal{C}}$ of each band is also given in the figures. The Berry phase of a single band varies between $0$ and $1$ for all the bands except a $3$ for $E_8$ and a $-2$ for $E_7$, which averages to be $1/Q$. The lowest band of conventional SkX has a sharp Dirac-cone-like structure, giving rise to a Berry phase of $Q=1$. The lowest band of $Q=2$ SkX has a hump pattern. By closer inspection, it does not have a sharp cone tip. As a result, the Berry phase of $E_{1}$ is zero. For both $Q=2$ and $Q=1$ SkX, $E_2$ and $E_3$ have energy overlap in a part of the Brillouin zone, conductivity does not show a plateau between them. In the gap between $E_3$ and $E_4$, the quantization number of ${\sigma}_{xy}$ is $3$ in the $Q=1$ SkX and $1$ in the $Q=2$ SkX. The intertwined $E_2$ and $E_3$ band group has a total Berry phase of $2$ in the $Q=1$ SkX and $1$ in the $Q=2$ SkX. This means that in average each band contributes a unitary Berry phase in the $Q=1$ SkX and each band contributes a $1/Q$ Berry phase in the $Q=2$ SkX. By comparison of the band profile between the two figures, we see that although the varying tendency of $E_2$ and $E_3$ is similar in the two types of SkX, the minimum valley of $Q=2$ SkX is smoother than that of the $Q=1$ SkX giving rise to a smaller Berry curvature. The band profile of $E_4$ in the $Q=2$ SkX is mirror symmetric with that of the $Q=1$ SkX. Therefore their Berry curvature and Berry phase are identical to each other. In the $Q=2$ SkX, bands $E_5$ to $E_{10}$ forms a group with any two adjacent bands overlapping each other vertically and the total Berry phase of the group is $6/Q$. Each band contributes a Berry phase of $1/Q$ in average. Contrastingly, in the $Q=1$ SkX, band $E_5$ and $E_9$ each has a unitary Berry phase; bands $E_6$ to $E_8$ form an overlapping group and have a Berry phase of $3Q$ in total. In the $Q=1$ SkX, bands $E_{10}$ and $E_{11}$ are overlapped to each other and the Hall conductivity does not present a plateau above $E_{10}$. Difference of the profile of bands $E_5$ to $E_{10}$ between the two types of SkX is also visible by comparing the two figures. Surface of the bands in the $Q=2$ SkX has more fluctuations in the momentum space than that in the $Q=1$ SkX. Usually this phenomenon accompany more trivial topology giving rise to smaller Berry curvature and smaller Berry phase, which is the case shown in the Hall conductivity. From the quantization number of the Hall conductivity in all the bands, we can see that each band of the $Q=2$ SkX has a Berry phase of $1/Q$ in average, which is remarkably different from conventional SkX with each band homogeneously bearing a unitary Berry phase.

Berry phase of the spin field of the SkX has two origins. One is the topology of each skyrmion. The other is the topology of the crystal. It has already been found that the $Q=1$ square-lattice\cite{HamamotoPRB2015} SkX has a topological Hall conductivity with steps of $1 \cdot  e^2/h$ and the $Q=1$ triangular-lattice\cite{GobelPRB2017} SkX has a topological Hall conductivity with steps of $2 \cdot e^2/h$ below and
with steps of $1 \cdot e^2/h$ above the van Hove singularity. The difference in the Hall conductivity originates from the topological difference in the two types of crystal structure. In this work, we consider the topological Hall effect of the $Q=2$ square-lattice SkX and found that each band contributes a Berry phase of $1/Q$. The case of the $Q=1$ SkX is a special one because of the identity $1=Q=1/Q$. If we take the step of $1 \cdot  e^2/h$ in the topological Hall conductivity to be $1/Q \cdot e^2/h$, results of the $Q=2$ SkX bear similar properties to the $Q=1$ SkX. Because the topological number of a single skyrmion is the Berry phase measured in the real space, it is not against intuition that the Berry phase becomes $1/Q$ in the momentum space. Because the momentum space of a single skyrmion is not well defined, the topological Hall conductivity is defined in a crystal with topologically-nontrivial band structure and a single skyrmion does not sustain a definite Hall conductivity. Topological properties of high-topological-number skyrmions are demonstrated in the crystal of skyrmions.

\begin{table}[h!]
  \begin{center}
    \caption{Berry phase of the SkX with $Q=3$ and $9 \times 9$ sublattices }
    \label{TabelQ3BerryPhase}
    \rule{16.04 cm}{0.05 cm}
    \begin{tabular}{c|c|c|c|c|c|c|c|c|c|c|c|c|c|c|c|c|c|c}
    \hline
     Band No.& $\bf{E_1}$ & $\bf{E_2}$& $\bf{E_3}$ & $\bf{E_4}$& $\bf{E_5}$ & $\bf{E_6}$& $\bf{E_7}$ & $\bf{E_8}$& $\bf{E_9}$ & $\bf{E_{10}}$& $\bf{E_{11}}$ & $\bf{E_{12}}$& $\bf{E_{13}}$& $\bf{E_{14}}$ & $\bf{E_{15}}$& $\bf{E_{16}}$ & $\bf{E_{17}}$& $\bf{E_{18}}$\\
            \hline
      $\cal{C}$ & 0& 1& 0 & 0& 2 & -1& 0 & 0& 2 & -2& 1 & 2& -1 & 0& 2 & -1& 0 & 1\\
      \hline
      \emph{avg}. $ \cal{C}$ & \multicolumn{3}{c|}{1/3} & \multicolumn{3}{c|}{1/3}& \multicolumn{3}{c|}{2/3}& \multicolumn{3}{c|}{1/3}& \multicolumn{3}{c|}{1/3}& \multicolumn{3}{c}{0}
      \\
      \hline
     Band No.& $\bf{E_{19}}$ & $\bf{E_{20}}$& $\bf{E_{21}}$ & $\bf{E_{22}}$& $\bf{E_{23}}$ & $\bf{E_{24}}$& $\bf{E_{25}}$ & $\bf{E_{26}}$& $\bf{E_{27}}$ & $\bf{E_{28}}$& $\bf{E_{29}}$ & $\bf{E_{30}}$\\
            \hline
      $\cal{C}$ & 0& 1& 0 & -2& 3 & 0 & 0 & 0& 1 & 0& 3 & -2\\
      \hline
      \emph{avg}. $ \cal{C}$ & \multicolumn{3}{c|}{1/3} & \multicolumn{3}{c|}{1/3}& \multicolumn{3}{c|}{1/3}& \multicolumn{3}{c|}{1/3}
      \\
      \hline
    \end{tabular}
    \rule{16.04 cm}{0.05 cm}
  \end{center}
\end{table}

To confirm the conclusion drawn from the numerical results of the SkX with $Q=2$, we conducted the computation in the SkX with $Q=3$. Because the skyrmion with a higher topological number has a finer spin texture, to secure accuracy in the latter case we consider each skyrmion consisting $9 \times 9 =81$ atoms. Corresponding results are shown in Table \ref{TabelQ3BerryPhase}. From the table, we can see that except two cases all the 1 to 30 bands counting from the lowest reproduced the $1/Q$ rule of the $Q=2$ SkX. The deviation should originate from the coarse $9 \times 9$-atom sublattice in comparison with the continuous momentum space.

Now we are more confident about the discovery that in the high-topological-number SkX as well as in the conventional SkX the Berry phase of each band averages to be $1/Q$ and the quantized zero-temperature Hall conductivity increases with a step characterizing the band Berry phase. Although there are a small percentage of deviations in the result of $Q=3$, we assume them tolerable and the conclusion is trustable based on the following two facts. First, the topological number of a single skyrmion is calculated assuming a continuous spin field. Considerable deviation occurs if the lattice one used is not fine enough, e.g., an $10 \times 10$ lattice is enough to introduce several percent of deviation of $Q$ and the deviation is more prominent for larger $Q$. Second, we have compared the results of the Berry phase $\cal{C}$ or identically the zero-temperature Hall conductivity ${\sigma}_{xy}$ of the high-topological-number SkX among the $9 \times 9 $, $5 \times 5$, and $4 \times 4$ sublattices, the last of which is used in Ref. \onlinecite{HamamotoPRB2015}. And the higher the topological number $Q$ the larger the unit-cell size is required to have a satisfactory result because the skyrmion with higher $Q$ bears a finer spin texture.

\section{Conclusions}

In the limit of large Hund's interaction, the free-electron system coupled with the background spin texture of the SkX can be approximated to a spinless ``tight-binding" model with the local hopping energy determined by the spin field of the SkX. We extend previous approaches to the Hall conductivity in the conventional SkX to the high-topological-number SkX with $Q=2$ and $Q=3$. We found that the Berry phase ${\cal{C}}$ of a single band quantized between $0$ to $\pm{3}$ for all the bands. The sequential $Q$ bands form a group, which totally contributes a Berry phase of unity. In this way, the Berry phase of each band averages to be $1/Q$ and the Hall conductivity increases with a step smaller than the conventional SkX. We attribute the result to the fact that the Berry phase ${\cal{C}}$ is measured in the momentum space and the quantum number of a single skyrmion $Q$ is measured in the real space. The reciprocality does not affect the conventional SkX because $Q=1=1/Q$.

\section{Acknowledgements}

We acknowledge support by the National Natural Science Foundation of China (No. 11004063) and the Fundamental Research Funds for the Central Universities, SCUT (No. 2017ZD099).

\section{Appendix: Equality of the zero-temperature Hall conductivity calculated from the Kubo formula to the total Berry phase of the bands below the Fermi energy}

From Eq. (\ref{sigma}),
\begin{equation}
\begin{array}{l}
\left\langle {n{\bf{k}}} \right|\frac{{\partial H}}{{\partial {k_x}}}\left| {m{\bf{k}}} \right\rangle  = \left\langle {n{\bf{k}}} \right|\left( {\frac{\partial }{{\partial {k_x}}}\left( {H\left| {m{\bf{k}}} \right\rangle } \right) - H\frac{{\partial \left| {m{\bf{k}}} \right\rangle }}{{\partial {k_x}}}} \right)\\
 = \left\langle {n{\bf{k}}} \right|\left( {\frac{\partial }{{\partial {k_x}}}\left( {{E_{m{\bf{k}}}}\left| {m{\bf{k}}} \right\rangle } \right) - H\frac{{\partial \left| {m{\bf{k}}} \right\rangle }}{{\partial {k_x}}}} \right)\\
 = \frac{{\partial {E_{m{\bf{k}}}}}}{{\partial {k_x}}}\left\langle {n{\bf{k}}} \right|\left. {m{\bf{k}}} \right\rangle  + {E_{m{\bf{k}}}}\left\langle {n{\bf{k}}} \right|\frac{\partial }{{\partial {k_x}}}\left| {m{\bf{k}}} \right\rangle  - {E_{n{\bf{k}}}}\left\langle {n{\bf{k}}} \right|\frac{\partial }{{\partial {k_x}}}\left| {m{\bf{k}}} \right\rangle .
\end{array}
\label{A1}
\end{equation}
Because $m \ne n$, $\left\langle {n{\bf{k}}} \right|\left. {m{\bf{k}}} \right\rangle  = 0$. Hence, we have
\begin{equation}
\begin{array}{l}
\sum\limits_{{E_n} < {E_F},m \ne n} {\frac{{\left\langle {n{\bf{k}}} \right|\frac{{\partial H}}{{\partial {k_x}}}\left| {m{\bf{k}}} \right\rangle \left\langle {m{\bf{k}}} \right|\frac{{\partial H}}{{\partial {k_y}}}\left| {n{\bf{k}}} \right\rangle  - \left( {n \leftrightarrow m} \right)}}{{{{\left( {{E_{n{\bf{k}}}} - {E_{m{\bf{k}}}}} \right)}^2}}}} \\
 = \sum\limits_{{E_n} < {E_F},m \ne n} {\frac{{ - {{\left( {{E_{m{\bf{k}}}} - {E_{n{\bf{k}}}}} \right)}^2}\left\langle {n{\bf{k}}} \right|\frac{\partial }{{\partial {k_x}}}\left| {m{\bf{k}}} \right\rangle \left\langle {m{\bf{k}}} \right|\frac{\partial }{{\partial {k_y}}}\left| {n{\bf{k}}} \right\rangle  - \left( {n \leftrightarrow m} \right)}}{{{{\left( {{E_{n{\bf{k}}}} - {E_{m{\bf{k}}}}} \right)}^2}}}} \\
 = \sum\limits_{{E_n} < {E_F},m \ne n} {\left( {\left\langle {m{\bf{k}}} \right|\frac{\partial }{{\partial {k_x}}}\left| {n{\bf{k}}} \right\rangle \left\langle {n{\bf{k}}} \right|\frac{\partial }{{\partial {k_y}}}\left| {m{\bf{k}}} \right\rangle  - \left\langle {n{\bf{k}}} \right|\frac{\partial }{{\partial {k_x}}}\left| {m{\bf{k}}} \right\rangle \left\langle {m{\bf{k}}} \right|\frac{\partial }{{\partial {k_y}}}\left| {n{\bf{k}}} \right\rangle } \right)} .
\end{array}
\end{equation}
Continuing on, we can have
\begin{equation}
\begin{array}{l}
\sum\limits_{{E_n} < {E_F},m \ne n} {\left( {\left\langle {m{\bf{k}}} \right|\frac{\partial }{{\partial {k_x}}}\left| {n{\bf{k}}} \right\rangle \left\langle {n{\bf{k}}} \right|\frac{\partial }{{\partial {k_y}}}\left| {m{\bf{k}}} \right\rangle  - \left\langle {n{\bf{k}}} \right|\frac{\partial }{{\partial {k_x}}}\left| {m{\bf{k}}} \right\rangle \left\langle {m{\bf{k}}} \right|\frac{\partial }{{\partial {k_y}}}\left| {n{\bf{k}}} \right\rangle } \right)} \\
 = \sum\limits_{{E_n} < {E_F},m \ne n} {\left( {\left\langle {n{\bf{k}}} \right|\frac{\partial }{{\partial {k_y}}}\left| {m{\bf{k}}} \right\rangle \left\langle {m{\bf{k}}} \right|\frac{\partial }{{\partial {k_x}}}\left| {n{\bf{k}}} \right\rangle  - \left\langle {n{\bf{k}}} \right|\frac{\partial }{{\partial {k_x}}}\left| {m{\bf{k}}} \right\rangle \left\langle {m{\bf{k}}} \right|\frac{\partial }{{\partial {k_y}}}\left| {n{\bf{k}}} \right\rangle } \right)} \\
 = \sum\limits_{{E_n} < {E_F},m \ne n} {\left( \begin{array}{l}
\left( {\frac{\partial }{{\partial {k_y}}}\left\langle {n{\bf{k}}} \right|\left. {m{\bf{k}}} \right\rangle  - \left( {\frac{\partial }{{\partial {k_y}}}\left\langle {n{\bf{k}}} \right|} \right)\left| {m{\bf{k}}} \right\rangle } \right)\left\langle {m{\bf{k}}} \right|\frac{\partial }{{\partial {k_x}}}\left| {n{\bf{k}}} \right\rangle \\
 - \left( {\frac{\partial }{{\partial {k_x}}}\left\langle {n{\bf{k}}} \right|\left. {m{\bf{k}}} \right\rangle  - \left( {\frac{\partial }{{\partial {k_x}}}\left\langle {n{\bf{k}}} \right|} \right)\left| {m{\bf{k}}} \right\rangle } \right)\left\langle {m{\bf{k}}} \right|\frac{\partial }{{\partial {k_y}}}\left| {n{\bf{k}}} \right\rangle
\end{array} \right)} \\
 = \sum\limits_{{E_n} < {E_F},m \ne n} {\left( {\left( { - \left( {\frac{\partial }{{\partial {k_y}}}\left\langle {n{\bf{k}}} \right|} \right)\left| {m{\bf{k}}} \right\rangle } \right)\left\langle {m{\bf{k}}} \right|\frac{\partial }{{\partial {k_x}}}\left| {n{\bf{k}}} \right\rangle  - \left( { - \left( {\frac{\partial }{{\partial {k_x}}}\left\langle {n{\bf{k}}} \right|} \right)\left| {m{\bf{k}}} \right\rangle } \right)\left\langle {m{\bf{k}}} \right|\frac{\partial }{{\partial {k_y}}}\left| {n{\bf{k}}} \right\rangle } \right)} .
\end{array}
\end{equation}
Using
\begin{equation}
\sum\limits_{m\left( {m \ne n} \right)} {\left| {m{\bf{k}}} \right\rangle \left\langle {m{\bf{k}}} \right|}  = 1 - \left| {n{\bf{k}}} \right\rangle \left\langle {n{\bf{k}}} \right|,
\end{equation}
we can have
\begin{equation}
\begin{array}{l}
\sum\limits_{{E_n} < {E_F},m \ne n} {\left( {\left( { - \left( {\frac{\partial }{{\partial {k_y}}}\left\langle {n{\bf{k}}} \right|} \right)\left| {m{\bf{k}}} \right\rangle } \right)\left\langle {m{\bf{k}}} \right|\frac{\partial }{{\partial {k_x}}}\left| {n{\bf{k}}} \right\rangle  - \left( { - \left( {\frac{\partial }{{\partial {k_x}}}\left\langle {n{\bf{k}}} \right|} \right)\left| {m{\bf{k}}} \right\rangle } \right)\left\langle {m{\bf{k}}} \right|\frac{\partial }{{\partial {k_y}}}\left| {n{\bf{k}}} \right\rangle } \right)} \\
 = \sum\limits_{{E_n} < {E_F}} {\left( { - \left( {\frac{\partial }{{\partial {k_y}}}\left\langle {n{\bf{k}}} \right|} \right)\frac{\partial }{{\partial {k_x}}}\left| {n{\bf{k}}} \right\rangle  + \left( {\frac{\partial }{{\partial {k_x}}}\left\langle {n{\bf{k}}} \right|} \right)\frac{\partial }{{\partial {k_y}}}\left| {n{\bf{k}}} \right\rangle } \right)} \\
 + \sum\limits_{{E_n} < {E_F}} {\left( {\left( {\left( {\frac{\partial }{{\partial {k_y}}}\left\langle {n{\bf{k}}} \right|} \right)\left| {n{\bf{k}}} \right\rangle } \right)\left\langle {n{\bf{k}}} \right|\frac{\partial }{{\partial {k_x}}}\left| {n{\bf{k}}} \right\rangle  - \left( {\left( {\frac{\partial }{{\partial {k_x}}}\left\langle {n{\bf{k}}} \right|} \right)\left| {n{\bf{k}}} \right\rangle } \right)\left\langle {n{\bf{k}}} \right|\frac{\partial }{{\partial {k_y}}}\left| {n{\bf{k}}} \right\rangle } \right)} .
\end{array}
\label{A2}
\end{equation}
The second term on the right hand side of Eq. (\ref{A2}) is
\begin{equation}
\begin{array}{l}
\sum\limits_{n < {E_F}} {\left( {\left( {\left( {\frac{\partial }{{\partial {k_y}}}\left\langle {n{\bf{k}}} \right|} \right)\left| {n{\bf{k}}} \right\rangle } \right)\left\langle {n{\bf{k}}} \right|\frac{\partial }{{\partial {k_x}}}\left| {n{\bf{k}}} \right\rangle  - \left( {\left( {\frac{\partial }{{\partial {k_x}}}\left\langle {n{\bf{k}}} \right|} \right)\left| {n{\bf{k}}} \right\rangle } \right)\left\langle {n{\bf{k}}} \right|\frac{\partial }{{\partial {k_y}}}\left| {n{\bf{k}}} \right\rangle } \right)} \\
 = \sum\limits_{n < {E_F}} {\left( {\left\langle {n{\bf{k}}} \right|\frac{\partial }{{\partial {k_x}}}\left| {n{\bf{k}}} \right\rangle \left( {\frac{\partial }{{\partial {k_y}}}\left\langle {n{\bf{k}}} \right|} \right)\left| {n{\bf{k}}} \right\rangle  - \left\langle {n{\bf{k}}} \right|\frac{\partial }{{\partial {k_y}}}\left| {n{\bf{k}}} \right\rangle \left( {\frac{\partial }{{\partial {k_x}}}\left\langle {n{\bf{k}}} \right|} \right)\left| {n{\bf{k}}} \right\rangle } \right)} \\
 = \sum\limits_{n < {E_F}} {\left( {\left\langle {n{\bf{k}}} \right|\frac{\partial }{{\partial {k_x}}}\left| {n{\bf{k}}} \right\rangle \left( {\frac{\partial }{{\partial {k_y}}}\left\langle {n{\bf{k}}} \right|} \right)\left| {n{\bf{k}}} \right\rangle  - \left\langle {n{\bf{k}}} \right|{{\left( {\frac{\partial }{{\partial {k_x}}}\left| {n{\bf{k}}} \right\rangle \left( {\frac{\partial }{{\partial {k_y}}}\left\langle {n{\bf{k}}} \right|} \right)} \right)}^\dag }\left| {n{\bf{k}}} \right\rangle } \right)} \\
 = 0.
\end{array}
\end{equation}
It is equal to zero because the expectation value of the conjugate of any operator is equal to that of the operator itself. The first term of Eq. (\ref{A2}) is
\begin{equation}
\sum\limits_{{E_n} < {E_F}} {\left( {\left( {\frac{\partial }{{\partial {k_x}}}\left\langle {n{\bf{k}}} \right|} \right)\frac{\partial }{{\partial {k_y}}}\left| {n{\bf{k}}} \right\rangle  - \left( {\frac{\partial }{{\partial {k_y}}}\left\langle {n{\bf{k}}} \right|} \right)\frac{\partial }{{\partial {k_x}}}\left| {n{\bf{k}}} \right\rangle } \right)} ,
\end{equation}
which multiplied by $ - \frac{i}{{2\pi }}$ and integrated over the first Brillouin zone is just the total Berry phase of the bands below the Fermi energy defined by Eqs. (\ref{BerryPhase}) and (\ref{BerryCurvature}).

\clearpage

\begin{figure}[ht]
\includegraphics[height=10cm, width=14cm]{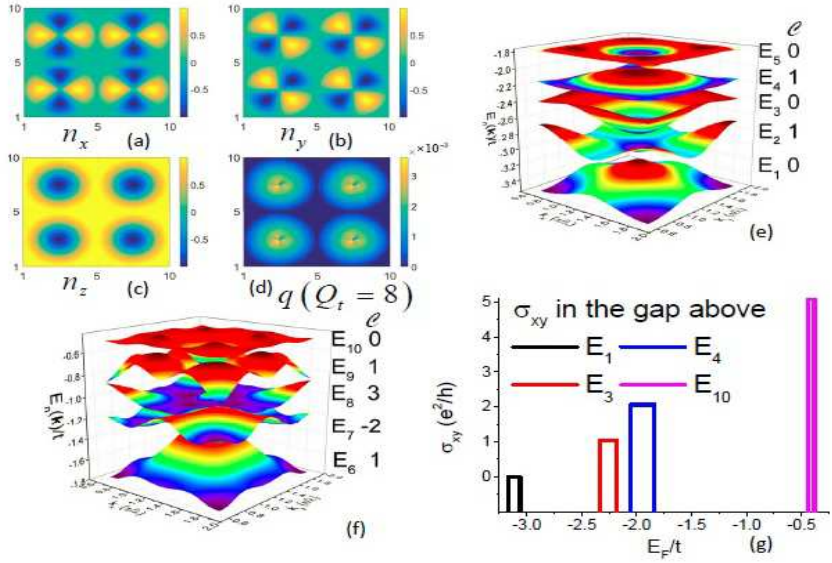}
\caption{ (a) (b) (c), and (d) Top views of the magnetization distributions $n_x$, $n_y$, and $n_z$ of the high-topological-number SkX, the corresponding topological charge density distribution $q$, respectively. Because each skyrmion has a topological number $Q=2$, the four-skyrmion array has the total topological number $Q_t=8$. (e) (f) Band structure of the lowest ten bands in the vicinity of the $M$ $({\pi}/{\lambda},{\pi}/{\lambda})$ point in the first Brillouin zone. Serial number of the bands is labeled by $E_1$ to $E_{10}$ counting from the lowest to the highest. The Berry phase ${\cal{C}}$ of each band is given beside its serial number. (g) Zero-temperature topological Hall conductivity as a function of the Fermi energy, which demonstrates plateaus in the energy gap between adjacent single band or intertwined multiple bands. It can be seen that averagely each band contributes a Berry phase of $1/Q$ to the topological Hall conductivity.
 }
\end{figure}
\clearpage
\begin{figure}[ht]
\includegraphics[height=10cm, width=14cm]{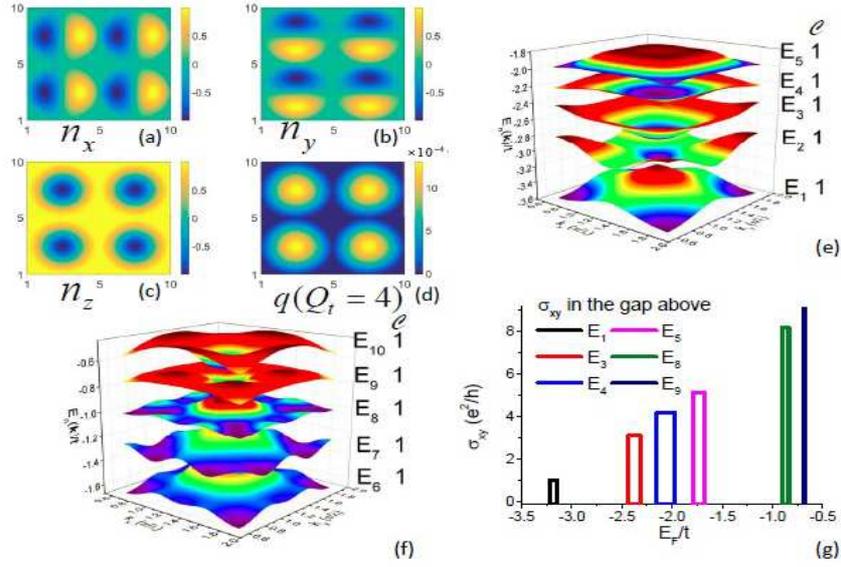}
\caption{(a) (b) (c), and (d) Top views of the magnetization distributions $n_x$, $n_y$, and $n_z$ of the conventional SkX, the corresponding topological charge density distribution $q$, respectively. Because each skyrmion has a topological number $Q=1$, the four-skyrmion array has the total topological number $Q_t=4$. (e) (f) Band structure of the lowest ten bands in the vicinity of the $M$ $({\pi}/{\lambda},{\pi}/{\lambda})$ point in the first Brillouin zone. Serial number of the bands is labeled by $E_1$ to $E_{10}$ counting from the lowest to the highest. The Berry phase ${\cal{C}}$ of each band is given beside its serial number. (g) Zero-temperature topological Hall conductivity as a function of the Fermi energy, which demonstrates plateaus in the energy gap between adjacent single band or intertwined multiple bands. It can be seen that each band contributes a Berry phase of unity to the topological Hall conductivity. }
\end{figure}
\clearpage

\begin{figure}[ht]
\includegraphics[height=5cm, width=6cm]{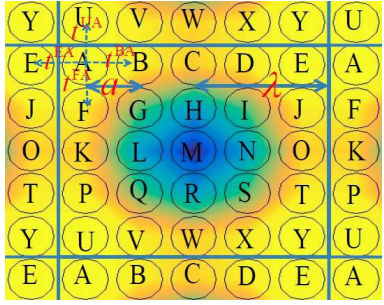}
\caption{Schematics of the ``tight-binding" model on the square-lattice SkX. We consider the radius of the skyrmion $\lambda =2.5a$. Each skyrmion can be taken as a giant unit cell consisting of $5 \times 5 = 25$ atoms labeled by capital letters A to Y. Background is top view of the magnetization distribution $n_z$ of the high-topological-number SkX.  }
\end{figure}
\clearpage

\end{document}